\documentstyle[aps,manuscript]{revtex}

\newcommand{\be}{\begin{equation}}
\newcommand{\ee}{\end{equation}}

\def\fun#1#2{\lower3.6pt\vbox{\baselineskip0pt\lineskip.9pt
\ialign{$\mathsurround=0pt#1\hfil##\hfil$\crcr#2\crcr\sim\crcr}}}

\begin{document}

\def\baselinestretch{1.5}
\normalsize

\hspace*{7cm}{1998/28 IKP-Theorie}

\title{The Correlator of Topological Charge Densities\\
at low $Q^2$ in QCD.}
\author{B. L. Ioffe\footnote{e-mail: ioffe@vxitep5.itep.ru}}
\address{Institute of Theoretical
and Experimental Physics,\\
 B.Cheremushkinskaya 25,
 Moscow 117218, Russia\footnote{permanent address}\\
 and\\
 Institut f\"ur Kernphysik, FZ J\"ulich,\\
 52425, J\"ulich, Germany.}

\maketitle

\begin{abstract}
The correlator of topological charge densities $\chi(Q^2)$ in QCD is
calculated in the domain of low $Q^2$. Basing on the Ward identities the low
energy theorems are proved. The value of the first derivative
$\chi^{\prime}(0)$ found recently by QCD sum rules \cite{1} is used. The
contributions of pseudoscalar quasi-Goldstone bosons as intermediate states
in the correlator are calculated with the account of $\eta-\pi$ mixing.

PACS number(s): 12.38. Aw, 11.55, 11.15. Tk, 11.30. Rd

\end{abstract}

\def\baselinestretch{1.5}
\normalsize
\vspace{2mm}
\vspace{2mm}

\section{Introduction}

The  existence of topological quantum number is a very specific feature of
non-abelian quantum field theories and, particularly, QCD. Therefore, the
study of properties of the topological charge density operator in QCD

\be
Q_5(x)= \frac{\alpha_s}{8\pi} G^n_{\mu \nu}(x)\tilde{G}^n_{\mu\nu}(x)
\ee
and of the corresponding vacuum correlator

\be
\chi(q^2) = i\int d^4x e^{iqx} \langle 0\mid T\left \{ Q_5(x),~Q_5(0)\right
\}\mid 0 \rangle
\ee
is of a great theoretical interest. (Here $G^n_{\mu\nu}$ is gluonic field
strength tensor, $\tilde{G}_{\mu\nu}=$
$(1/2)\varepsilon_{\mu\nu\lambda\sigma} G_{\lambda \sigma}$ is its dual, $n$
are
the colour indeces, $n=1,2,... N^2_c-1$, $N_c$ is the number of colours,
$N_c=3$ in QCD). The existence of topological quantum numbers in non-abelian
field theories was first discovered by Belavin et al.\cite{2}, their connection
with non-conservation of $U(1)$ chirality was established by t'Hooft
\cite{3}. Crewther \cite{4}  derived Ward identities related to $\chi(0)$,
which allowed him to prove the theorem, that $\chi(0)=0$  in any theory
where it is at least one massless quark. An important step in the
investigation of the properties of $\chi(q^2)$ was achieved by Veneziano
\cite{5} and Di Vecchia and Veneziano \cite{6}. These authors considered the
limit $ ~N_c\to \infty$. Assuming that in the theory there are $N_f$ light
quarks with the masses $m_i \ll M$, where $M$ is the characteristic scale of
strong interaction, Di Vecchia and Veneziano found that

\be
\chi(0) = \langle 0\mid \bar{q}q\mid 0\rangle \Biggl (
\sum^{N_f}_i~\frac{1}{m_i}\Biggr )^{-1},
\ee
where $\langle 0\mid \bar{q}q\mid 0\rangle $ is the common value of quark
condensate for all light quarks and the terms of the order $m_i/M$ are
neglected. \footnote{The definition of $\chi(q^2)$ used  above eq.2,
differs by sign from the definition used in \cite{4}-\cite{6}.} The concept
of $\theta$-term in the Lagrangian was succesfully exploited in \cite{6} in
deriving of (3). Using the same concept and studying the properties of the
Dirac operator Leutwyler and Smilga \cite{7}  succeeded in proving eq.3 at
any $N_c$ for the case of two light quarks, $u$ and $d$.

In this paper I calculate $\chi(q^2)$, at low $\mid q^2 \mid \ll M^2$. In
ref.1 the value $\chi^{\prime}(0)$ (more precise, its nonperturbative part)
was found. On the basis of QCD sum rules in the external fields the
connection of $\chi^{\prime}(0)$  with the part of the
proton spin, carried by $u,d,s $  quarks, $\Sigma$  was established. By
the use of experimental data on $\Sigma$, as well as from the requirement of
selfconsistency of the sum rule, it was obtained:

\be
\chi^{\prime}(0)=(2.3 \pm 0.6) \times 10^{-3}~GeV^2
\ee
in the limit of massless $u,d$ and $s$ quarks. At low $q^2$ the terms,
proportional to quark masses, are related to the contributions of light
pseudoscalar mesons as intermediate states in the correlator (2). These
contributions are calculated below.  In such calculation for the case of
three quarks the mixing of $\pi^0$ and $\eta$  is of importance and it is
accounted.

The presentation of the material in the paper is the following. In Sec.II the
low energy theorems related to $\chi(0)$  are rederived with the account of
possible anomalous equal-time commutator terms. (In \cite{4}-\cite{6} it was
implicitly assumed that these terms are zero). In Sec.III the case of one and
two light quarks are considered. It is proved, that the mentioned above
commutator terms are zero indeed and for the case of two quarks eq.3 is
reproduced without using $N_c \to \infty$ limit and the concepts of
$\theta$--terms.  In   Sec.IV the case of three $u,d,s$ light quarks is
considered in the approximation $m_u,m_d \ll m_s$. The problem of mixing of
$\pi^0$  and $\eta$  states \cite{8,9} is formulated and corresponding
 formulae are
presented. The account of $\pi-\eta$ mixing allows one to get eq.(3) from
low energy theorems, formulated in Sec.II for the case of three light quarks.
(At $m_u, m_d \ll m_s$ it coincides with the two quark case). In Sec.V the
$q^2$--dependence of $\chi(q^2)$ was found at low $\mid q^2 \mid$ in the
leading nonvanishing order in $q^2/ M^2$ as well as in $m_q/M$.

\section{Low energy theorems.}

Consider QCD with $N_f$  light quarks, $m_i \ll M \sim 1~GeV$, $i=1,...
N_f$. Define the singlet (in flavour) axial current by

\be
j_{\mu 5}(x) = \sum^{N_f}_i \bar{q}_i(x)\gamma_{\mu}\gamma_5 q(x)
\ee
and the polarization operator

\be
P_{\mu\nu} (q) = i\int d^4x e^{iqx} \langle 0\mid T~ \{j_{\mu 5}(x),
j_{\mu 5}(0)~\} \mid 0 \rangle.
\ee
The general form of the polarization operator is:

\be
P_{\mu \nu} (q) = -P_L(q^2)\delta_{\mu\nu} + P_T (q^2)(-\delta_{\mu\nu} q^2
+ q_{\mu}q_{\nu})
\ee
Because of anomaly the singlet axial current is nonconserving:

\be
\partial_{\mu}j_{\mu 5}(x) = 2N_f Q_5(x) + D(x),
\ee
where $Q_5(x)$ is given by (1) and

\be
D(x) = 2i\sum^{N_f}_i m_i\bar{q}_i (x)\gamma_5 q_i(x)
\ee
It is well known, that even if some light quarks are massless, the
corresponding Goldstone bosons, arising from spontaneous violation of chiral
symmetry do not contribute to singlet axial channel (it is the solution of
$U(1)$ problem), i.e. to polarization operator $P_{\mu\nu}(q)$. $P_L(q^2)$
also have no kinematical singularities at $q^2=0$. Therefore

\be
P_{\mu\nu} (q)q_{\mu}q_{\nu} = -P_L(q^2)q^2
\ee
vanishes in the limit $q^2 \to 0$. Calculate the left-hand side (lhs) of (10)
in the standard way -- put $q_{\mu}q_{\nu}$ inside the integral in (6) and
integrate by parts. (For this it is convenient to represent the polarization
operator in the coordinate space as a function of two coordinates $x$ and
$y$.) Going to the limit $q^2 \to 0$ we have

$$
\lim_{q^2\to 0}P_{\mu\nu} (q)q_{\mu}q_{\nu} = i\int d^4 x \langle 0 \mid T\{
2N_fQ_5(x),~2N_fQ_5(0)  $$
$$+ 2N_fQ_5(x),~D(0) + D(x), ~2N_fQ_5(0) + D(x),~D(0)\} \mid 0 \rangle $$
$$+4\sum^{N_f}_i m_i \langle \mid \bar{q}_i(0)q_i(0)\mid 0 \rangle  $$
\be
+\int d^4 x \langle  0 \mid ~[j_{05}(x),~2N_f Q_5 (0)~] \mid 0 \rangle
\delta(x_0)= 0
\ee
In the calculation of (11) the anomaly condition (8) was used. The terms,
proportional to quark condensates arise from equal time commutator
$[j_{05}(x)$, $D(0)]_{x_0=0}$, calculated by standard commutation relations.
Relation (11) up to the last term was first obtained by Crewther \cite{4}.
The last term, equal to zero according to standard commutation  relations
and omitted in \cite{4}-\cite{6}, is keeped. The reason is, that we deal
with very subtle situation, related to anomaly, where nonstandard Schwinger
terms in commutation relations may appear. (It can be shown, that, in
general the only Schwinger term in this problem is given by the last term in
the lhs of (11): no others can arise.) Consider also the correlator:

\be
P_{\mu}(q) = i\int d^4x~e^{iqx} \langle 0 \mid T \{j_{\mu 5}(x),~Q_5(0)\}
\mid 0 \rangle
\ee
and the product $P_{\mu}(q)q_{\mu}$  in the limit $q^2\to 0$ (or $q^2$ of
order of the $m^2_{\pi}$, where $m_{\pi}$ is the mass of Goldstone boson).
The general form of $P_{\mu}(q)$ is $P_{\mu}(q)=Aq_{\mu}$. Therefore
nonvanishing values of $P_{\mu}q_{\mu}$  in the limit $q^2 \to 0$ (or of
order of quark mass $m$, if $\mid q^2 \mid \sim m^2_{\pi}$ -- this limit will
be
also intersting for us later) can arise only from Goldstone bosons
intermediate states in (12).
Let us us estimate the corresponding matrix elements

\be
\langle 0 \mid j_{\mu 5}\mid \pi \rangle = F q_{\mu}
\ee
\be
\langle 0 \mid Q_5\mid \pi \rangle = F^{\prime}
\ee
$F$ is of order of $m$, since in the limit of massless quarks Goldstone
bosons are coupled only to nonsinglet axial current. $F^{\prime}$ is of order
of $m^2_{\pi}f_{\pi}\sim m$, where $f_{\pi}$ is the pion decay constant (not
considered to be small), since in massless quark limit, the Goldstone boson is
decoupled
 from $Q_5$ . These estimations give

\be
P_{\mu}q_{\mu} \sim \frac{q^2}{q^2 - m^2_{\pi}} m^2
\ee
and it is zero at $q^2 \to 0,~m^2_{\pi}\not=0$ and of order of $ m^2$ at
$q^2\sim
m^2_{\pi}$. In what follows I will restrict myself by the terms linear in
quark masses.  So, I can put $P_{\mu}(q)q_{\mu}=0$  at $q\to 0$. The
integration by parts, in the right-hand side (rhs) of (12) gives:

$$
\lim_{q^2\to 0} P_{\mu}(q)q_{\mu} = -\int d^4x \langle 0 \mid T \{ 2N_f
Q_5(x),~ Q_5(0) + D(x),~Q_5(0)\}\mid 0 \rangle $$
\be
-\int d^4 x \langle 0 \mid [~j_{05}(x),~Q_5(0)~]\mid 0 \rangle \delta(x_0) =0
\ee
After the substitution of (16) in (11) arise the low energy theorem:

$$i\int d^4 x \langle 0 \mid T \{ 2N_f Q_5(x),~ 2N_fQ_5(0)\} \mid 0\rangle
$$
$$-i\int d^4 x \langle 0\mid T\{ D(x),~D(0)\} \mid 0 \rangle - 4\sum^{N_f}_i
m_i \langle 0 \mid \bar{q}_i(0)q_i(0)\mid 0 \rangle $$
\be
+ i\int d^4 x \langle 0\mid  [~j^0_{05}(x),~2N_fQ_5(0)~] \mid 0\rangle
\delta(x_0) =0
\ee
The low energy theorem (17), with the last term in the lhs omitted, was
found by Crewther \cite{4}.

\section{One and two light quarks.}

Consider first the case of one massless quark, $N_f=1$, $m=0$. This case can
easily be treated by introduction of $\theta$-term in the Lagrangian,

\be
\Delta L = \theta \frac{\alpha_s}{4\pi} ~G^n_{\mu\nu}\tilde{G}^n_{\mu\nu}
\ee
The matrix element $\langle 0 \mid Q_5 \mid n \rangle$ between any hadronic
state $\mid n \rangle$ and vacuum is proportional

\be
\int d^4x \langle 0 \mid Q_5(x) \mid n \rangle  \sim \langle  0 \mid
\frac{\partial}{\partial\theta} lnZ\mid n \rangle_{\theta=0},
\ee
where $Z=e^{iL}$ and $L$ is the Lagrangian. The gauge transformation of the
quark field $\psi^{\prime}\to e^{i\alpha\gamma_5 }\psi$  results to appearance
of
the term

\be
\delta L = \alpha [~\partial_{\mu}j_{\mu 5} - (\alpha_s/4\pi)G^n_{\mu\nu}
\tilde{G}^n_{\mu\nu}~]
\ee
in the Lagrangian. By the choice $\alpha=\theta$ the $\theta$-term (18) will be
killed and $(\partial/\partial \theta)lnZ=0$. Therefore, $\chi(0)=0$
(Crewther theorem). The first term in (17) vanishes, as well as the second
and third, since $m=0$. From (17) we have, that indeed the anomalous
commutator vanishes

\be
\langle 0 \mid [~j_{05}(x),~Q_5(0)~]_{x_0=0} \mid 0 \rangle =0,
\ee
supporting the assumptions done in \cite{4}-\cite{6}.

Let us turn  now to the case of two light quarks, $u,d,N_f=2$. This is the
case of real QCD, where the strange quark is considered as a heavy. Define
the isovector axial current

\be
j^{(3)}_{\mu 5} = (\bar{u}\gamma_{\mu}\gamma_5 u -\bar{d}\gamma_{\mu}
\gamma_5 d)/\sqrt{2}
\ee
and its matrix element between the states of pion and vacuum

\be
\langle 0 \mid j^{(3)}_{\mu 5} \mid \pi \rangle = f_{\pi} q_{\mu},
\ee
where $q_{\mu}$ is pion 4-momentum, $f_{\pi}=133~MeV$. Multiply (23) by
$q_{\mu}$. Using Dirac equations for quark fields, we have

\newpage
$$\frac{2i}{\sqrt{2}}\langle  0\mid m_u
\bar{u}\gamma_{5}u -
m_d\bar{d}\gamma_{5}d \mid \pi \rangle
= \frac{i}{\sqrt{2}}
\langle  0\mid (m_u + m_d)(\bar{u}\gamma_{5}u - \bar{d}\gamma_{5}d) $$
\be
+ (m_u + m_d)((\bar{u}\gamma_{5}u - \bar{d}\gamma_{5}d)
\mid \pi \rangle = f_{\pi} m^2_{\pi},
\ee
where $m_u$, $m_d$  are $u$ and $d$ quark masses. The ratio of the matrix
elements in lhs  of (24) is of order

\be
\frac{\langle 0 \mid \bar{u}\gamma_{5}u +
\bar{d}\gamma_{5}d \mid \pi \rangle }
{\langle 0 \mid \bar{u}\gamma_{5}u -
\bar{d}\gamma_{5}d \mid \pi \rangle } \sim \frac{m_u-m_d}{M},
\ee
since the matrix element in the numerator violates isospin and this violaton
(in
the absence of elecntomagnetism, which is assumed)  can arise from the
difference $m_u-m_d$ only. Neglecting this matrix element we have from (24)

\be
\frac{i}{\sqrt{2}}\langle 0 \mid
\bar{u}\gamma_{5}u -
\bar{d}\gamma_{5}d
\mid \pi \rangle  = \frac{f_{\pi}m^2_{\pi}}{m_u+m_d}
\ee
Let us find $\chi(0)$ from low energy sum rule (17) restricting ourself to
the terms linear in quark masses. Since $D(x)\sim m$, the only intermediate
state contributing to  the matrix element

\be
\int d^4 x \langle  0 \mid T\{D(x),~ D(0)\} \mid 0 \rangle
\ee
in (17) is the one-pion state. Define

\be
D_q = 2i(m_u\bar{u}\gamma_5 u + m_d \bar{d}\gamma_5 d).
\ee
Then

$$\langle 0 \mid D_q \mid \pi \rangle = i\langle 0 \mid (m_u+m_d)
(\bar{u}\gamma_{5}u + \bar{d}\gamma_{5}d) + (m_u-m_d)
(\bar{u}\gamma_{5}u - \bar{d}\gamma_{5}d) \mid \pi \rangle$$

\be
= \sqrt{2} \frac{m_u-m_d}{m_u+m_d} f_{\pi}m^2_{\pi},
\ee
where the matrix element of singlet axial current was neglected  and (26)
was used. The substitution of (29) into (27) gives
\newpage
$$i\int d^4 x e^{iqx} \langle 0 \mid T \{ D_q(x),~D_q(0)\} \mid 0
\rangle_{q\to 0}= \lim_{q\to 0} \left \{ -\frac{1}{q^2-m^2_{\pi}}2\Biggl (
\frac{m_u - m_d}{m_u+m_d}\Biggr )^2 f^2_{\pi} m^2_{\pi}\right \} $$
\be
= - 4 \frac{(m_u - m_d)^2}{m_u+m_d}\langle 0 \mid \bar{q}q \mid 0 \rangle
\ee
In the last equality in (30) Gell-Mann-Oakes-Renner relation [10]

\be
\langle 0 \mid \bar{q}q \mid 0 \rangle =
-\frac{1}{2}~\frac{f^2_{\pi}m^2_{\pi}}{m_u+m_d}
\ee
was substituted as well the $SU(2)$ equalities

\be
\langle 0 \mid \bar{u}u \mid 0 \rangle =
\langle 0 \mid \bar{d}d \mid 0 \rangle \equiv
\langle 0 \mid \bar{q}q \mid 0 \rangle.
\ee
>From (17) and (30) we finally get:

\be
\chi(0) = i\int d^4 x \langle 0 \mid T\{ Q_5(x),~Q_5(0) \} \mid 0 \rangle =
\frac{m_um_d}{m_u + m_d}\langle 0 \mid \bar{q}q \mid 0 \rangle
\ee
in concidence with eq.3.

 In a similar way matrix element
$\langle 0 \mid Q_5 \mid \pi \rangle$ can be found. Consider

\be
\langle 0 \mid j_{\mu 5} \mid \pi \rangle = F q_{\mu}
\ee
The estimation of $F$  gives

\be
F\sim \frac{m_u - m_d}{M} f_{\pi}
\ee
and after multiplying of (34) by $q_{\mu}$ the rhs of (34) can be
neglected.
In the lhs we have

\be
\langle 0\mid D_q\mid \pi \rangle + 2N_f \langle 0\mid Q_5\mid \pi \rangle
=0
\ee
The substitution of (29) into (36) results in

\be
\langle 0 \mid Q_5 \mid \pi \rangle = -\frac{1}{2\sqrt2}\frac{m_u -
m_d}{m_u+m_d}
f_{\pi}m^2_{\pi}
\ee
The relation of this type (with a wrong numerical coefficient) was found in
\cite{9}, the correct formula was presented in \cite{11}. From comparison
 of (33) and (37) it is clear, that it would be wrong to calculate $\chi(0)$
 by accounting only pions as intermediate states in the lhs of (33) -- the
constant
 terms, reflecting the necessity of subtraction terms in dispersion relation
and
 represented by proportional to quark condensate terms in (17) are
 extremingly important. The cancellation of Goldstone bosons pole terms and
 these constant terms results in the Crewther theorem -- the vanishing of
 $\chi(0)$, when one of the quark masses, e.g. $m_u$  is going to zero.

\section{Three light quarks.}

Let us dwell on the real QCD case of three light quarks, $u,d$ and $s$. Since
the ratios $m_u/m_s$, $m_d/m_s$ are small, less than 1/20, account them only
in the leading order. When the $u$ and $d$  quark masses $m_u$ and $m_d$
are not assumed to be equal, the quasi-Goldstone states $\pi^0$  and $\eta$
are no more states of pure isospin 1 and 0 correspondingly: in both of these
states persist admixture of other isospin \cite{8,9} proporional to
$m_u-m_d$.
(The violation of isospin by electromagnetic interaction is small in the
problem under investigation \cite{8} and can be neglected. The
$\eta^{\prime} - \eta$ mixing is also neglected.) In order to treat the
problem it is convenient to introduce pure isospin 1 and 0 pseudoscalar
meson fields $\varphi_3$ and $\varphi_8$ in $SU(3)$ octet and the
corresponding states $\mid P_3 \rangle$, $\mid P_8 \rangle$ \cite{8,11}.
Then in the $SU(3)$ limit
\be
\langle 0\mid j^{(3)}_{\mu 5} \mid P_3 \rangle = f_{\pi}q_{\mu}
\ee
\be
\langle 0\mid j^{(8)}_{\mu 5} \mid P_8 \rangle = f_{\pi}q_{\mu},
\ee
where

\be
j^{(8)}_{\mu 5} = (\bar{u}\gamma_{\mu}\gamma_5 u +
\bar{d}\gamma_{\mu}\gamma_5 d - 2\bar{s}\gamma_{\mu}\gamma_5 s)/\sqrt{6}
\ee
and $j^{(3)}_{\mu 5}$ is given by (22). The states $\mid P_3 \rangle$,
$\mid P_8 \rangle$ are not eigenstates of the Hamiltonian. In the free
Hamiltonian

\be
H = \frac{1}{2}\tilde{m}^2_{\pi} \varphi^2_3 +
\frac{1}{2}\tilde{m}^2_{\eta}\varphi^2_8
+ \langle P_8 \mid P_3 \rangle \varphi_3\varphi_8  + \mbox{kinetic terms}
\ee
the nondiagonal term $\sim \varphi_3\varphi_8$ is present (
$\tilde{m}^2_{\pi}$ and $\tilde{m}^2_{\eta}$ in (41) coincides with
$m^2_{\pi}$ and $m^2_{\eta}$ up to terms quadratic in  $\langle P_8 \mid P_3
\rangle$).
The nondiagonal term was calculated in \cite{8} on the basis of PCAC  and
current algebra (see also \cite{9})

\be
\langle P_8 \mid P_3 \rangle = \frac{1}{\sqrt{3}}m^2_{\pi}
\frac{m_u-m_d}{m_u+m_d}
\ee
The physical  $\pi$ and $\eta$ states arise after orthogonalization of the
Hamiltonian (41)

$$
\hspace*{-0.8cm}\mid \pi \rangle = cos \theta \mid P_3 \rangle - sin \theta
\mid P_8
\rangle$$
\be
\mid \eta\rangle = sin \theta \mid P_3 \rangle + cos \theta \mid P_8 \rangle
\ee
where the mixing angle $\theta$ is given by (at small $\theta$) \cite{8,9}:

\be
\theta = \frac{\langle  P_8 \mid P_3 \rangle}{m^2_{\eta} - m^2_{\pi}}\approx
\frac{\langle P_8 \mid P_3 \rangle}{m^2_{\eta}} =
\frac{1}{\sqrt{3}}~\frac{m^2_{\pi}}{m^2_{\eta}}~\frac{m_u-m_d}{m_u+m_d}
\ee
In terms of the fields $\varphi_3$  and $\varphi_8$ PCAC relations take the
form \cite{8};

\be
\partial_{\mu}j^{(3)}_{\mu 5} = f_{\pi}(m^2_{\pi}\varphi_3 +
\langle  P_8 \mid P_3 \rangle \varphi_8)
\ee
\be
\partial_{\mu}j^{(8)}_{\mu 5} = f_{\pi}(m^2_{\pi}\varphi_8 +
\langle  P_8 \mid P_3 \rangle \varphi_3)
\ee
Our goal now is to calculate the contribution of pseudoscalar octet states
to the second term in the lhs of (17). It is convenient to use the full set
of orthogonal states $\mid P_3 \rangle$, $\mid P_8 \rangle $ as the basis.
Use the notation

\be
D = D_q + D_s ~~~~~~~~D_s = 2im_s\bar{s}\gamma_5 s,
\ee
where $D_q$  is given by (28). The matrix element

\be
\langle 0 \mid D_q \mid P_3 \rangle = \sqrt{2} \frac{m_u-m_d}{m_u + m_d}
f_{\pi}m^2_{\pi}
\ee
can be found by the same argumentation, as that was used in the derivation
of (29). In order to find $\langle 0 \mid D_q \mid P_8\rangle$ take the matrix
element of eq.45 between vacuum and $\mid P_8\rangle$

\be
\langle 0 \mid \partial_{\mu}j^{(3)}_{\mu 5} \mid P_3 \rangle =
f_{\pi} \langle  P_8 \mid P_3 \rangle
\ee
The substitution in the lhs of (49) of the expression for
$\partial_{\mu}j^{(3)}_{\mu 5} $ through quark fields gives

$$\langle 0 \mid \bar{u}\gamma_5 u - \bar{d}\gamma_5 d \mid P_8 \rangle = -
\frac{m_u - m_d}{m_u + m_d}
\langle 0 \mid \bar{u}\gamma_5 u + \bar{d}\gamma_5 d \mid P_8 \rangle -$$
\be
- i\sqrt{\frac{2}{3}} f_{\pi}m^2_{\pi} \frac{m_u - m_d}{(m_u + m_d)^2}
\ee
In a similar way take matrix element of eq. (46) between $\langle 0 \mid$ and
$\mid P_8 \rangle$  and substitute into it (50). We get
\newpage
$$\frac{i}{\sqrt{6}}\langle 0 \mid \Biggl [
m_u+m_d-\frac{(m_u - m_d)^2}
{m_u + m_d}\Biggr ]
(\bar{u}\gamma_5 u + \bar{d}\gamma_5 d) - 4m_s
\bar{s}\gamma_5 s \mid P_8 \rangle$$
\be
=f^2_{\pi}m^2_{\eta} - \frac{1}{3}m^2_{\pi}f_{\pi}\Biggl (
\frac{m_u - m_d}{m_u + m_d}\Biggr )^2
\ee
As  follows from  $SU(3)$ symmetry of strong interaction

\be
\langle 0 \mid \bar{u}\gamma_5 u + \bar{d}\gamma_5 d \mid P_8 \rangle = -
\langle 0 \mid\bar{s}\gamma_5 s \mid P_8 \rangle
\ee
up to terms of order $m_q/M$, which are neglected. From (51),(52) we find:

\be
\langle 0 \mid D_s\mid P_8 \rangle  = -\sqrt{\frac{3}{2}}f_{\pi}m^2_{\eta}
\Biggl [ 1 - \frac{1}{4} \frac{(m_u - m_d)^2}{m_s(m_u + m_d)}\Biggr ]
\Biggl [ 1 + \frac{m_u m_d}{m_s(m_u + m_d)}\Biggr ]
\ee

\be
\langle 0 \mid D_q\mid P_8 \rangle  = 4\sqrt{\frac{2}{3}}f_{\pi}m^2_{\pi}
\frac{m_u m_d}{(m_u + m_d)^2}
\ee
in notation (28),(47). When deriving (53) the $SU(3)$ relation

\be
m^2_{\eta} = \frac{4}{3} m^2_{\pi} \frac{m_s}{m_u+m_d}
\Biggl ( 1 - \frac{1}{4} \frac{m_u + m_d}{m_s}\Biggr )
\ee
was used. In (53) the small terms $\sim m_u/m_s$, $m_d/m_s$ are accounted,
because they are multiplyed by large factor
$m^2_{\eta}$. In (54) small terms are disregarded. The matrix element
$\langle 0 \mid D_s\mid P_3 \rangle $ can be found from (46). We have

\be
\frac{1}{\sqrt{6}}
\langle 0 \mid D_q - 2D_s \mid P_3 \rangle  = f_{\pi}
\langle P_8 \mid  P_3 \rangle
\ee
The substitution of (48) and (42) into (56) gives

\be
\langle 0 \mid D_s \mid P_3 \rangle  = 0
\ee
Equations (48),(53),(54)  and (57) allow one to calculate the interesting
for us correlator

\be
i \int d^4 x \langle 0 \mid T \{ D(x),~D(0) \} \mid 0 \rangle
\ee
when the sets $\mid P_3 \rangle \langle P_3 \mid$  and
$\mid P_8 \rangle \langle P_8 \mid$ are taken as intermediate states. But
$\mid P_3 \rangle,~ \mid  P_8  \rangle$ are not the eigenstates  of the
Hamiltonian, they mix in accord with (41). Therefore the transitions
$\langle  P_8 \mid P_3 \rangle$ arising from the mixing term in (41) must be
also accounted. There are two such terms. The one corresponds to the
transition $\langle 0\mid  D_s \mid P_8 \rangle$
$\langle P_3 \mid  D_q \mid 0 \rangle$ and its contribution to (58) is given
by

$$\lim_{q^2\to 0} \left \{ -2 \langle 0\mid  D_s \mid P_8 \rangle
\frac{1}{q^2 - m^2_{\eta}}\langle P_8 \mid P_3 \rangle
\frac{1}{q^2 - m^2_{\pi}} \langle P_3\mid  D_q \mid 0 \rangle \right \} =$$
\be
= 2f^2_{\pi} m^2_{\pi} \Biggl ( \frac{m_u - m_d}{m_u+m_d}\Biggr )^2
\ee
The other corresponds to the transition between two $D_s$ operators,
where $\langle P_3 \mid P_3 \rangle$ enter as intermediate state. This
contribution is equal to:

$$\lim_{q^2\to 0} \left \{ - \langle 0\mid  D_s \mid P_8 \rangle
\frac{1}{q^2 - m^2_{\eta}}\langle P_8 \mid P_3 \rangle
\frac{1}{q^2 - m^2_{\pi}} \langle P_8 \mid P_3 \rangle
\frac{1}{q^2 - m^2_{\eta}}
\langle P_8\mid  D_s \mid 0 \rangle \right \} =$$
\be
= \frac{1}{2}f^2_{\pi} m^2_{\pi} \Biggl (\frac{m_u - m_d}{m_u+m_d}\Biggr )^2
\ee
It is enough to account only matrix elements, with $D_s$ operators, since
they are enhanced by large factor $m^2_{\eta}$. All others are small in the
ratio $m^2_{\pi}/m^2_{\eta}$.

Collecting all together, we get:

$$i\int d^4 x \langle 0 \mid T \{ D(x),~D(0) \} \mid 0 \rangle=
f^2_{\pi} m^2_{\pi} \left \{ 2 \Biggl ( \frac{m_u -m_d}{m_u +m_d} \Biggr
)^2 - 8 \frac{m_u m_d}{(m_u +m_d)^2} \right. $$
$$ + 2\frac{m_s}{m_u +m_d}\Biggl ( 1 + \frac{1}{4}\frac{m_u +m_d}{m_s}
\Biggr ) \Biggl [ 1 - \frac{1}{2}\frac{m_u -m_d}{m_s(m_u+ m_d)} \Biggr ]
\Biggl [ 1 - 2\frac{m_u m_d}{m_s(m_u+ m_d)} \Biggr ]$$
\be
\left. + 2  \Biggl ( \frac{m_u -m_d}{m_u +m_d} \Biggr )^2 +
\frac{1}{2}\Biggl ( \frac{m_u -m_d}{m_u+ m_d} \Biggr )^2 \right \}=
f^2_{\pi} m^2_{\pi} \Biggl [  \frac{2m_s}{m_u +m_d}
+ \frac{9}{2}\Biggl ( \frac{m_u -m_d}{m_u+ m_d} \Biggr )^2 -
\frac{5}{2}\Biggr ]
\ee
The first term in the figure bracket in (61) comes from
$\langle 0\mid \ D_q\mid P_3\rangle^2$, the second -- from
$\langle 0\mid D_q\mid P_8\rangle \times \langle 0\mid P_8\mid D_s\rangle$, the
third --
from $\langle 0\mid D_s\mid P_8\rangle^2$, the last two terms are from
(59),(60).
Adding to (61) the proportional to quark condensate term

\be
4(m_u + m_d + m_s)\langle 0 \mid \bar{q}q \mid 0 \rangle
\ee
in (17), we finally get for 3 quarks at $m_u, m_d \ll m_s$

\be
i\int d^4 x \langle 0 \mid T \{ 2N_f Q_5(x),~ 2N_f Q_5(x)\} \mid 0 \rangle=
36 \frac{m_u m_d}{m_u +m_d} \langle 0 \mid \bar{q}q \mid 0 \rangle
\ee
and

\be
\chi(0) = \frac{m_u m_d}{m_u +m_d} \langle 0 \mid \bar{q}q \mid 0 \rangle,
\ee
since in this case $4 N_f^2=36$. Eq.64 coincides with eq.3, obtained \cite{6}
in $N_c
\to \infty~$ limit.
This fact demonstrates, that $N_c \to \infty~$ limit is irrelevant for
determination of $\chi(0)$ (at least for the cases of two or three light
quarks and at $m_u, m_d \ll m_s$). $\chi(0)$ for three light quarks at
$m_u,m_d \ll m_s$ -- eq.64 coincides with $\chi(0)$ in the two light quark
case, (see [7] and [33]) i.e in this problem, when $m_u,m_d \ll
m_s$, there is no difference if $s$-quark is considered as a heavy or light --
it softly appears in the theory.

Determine the matrix elements $\langle 0 \mid Q_5\mid \eta \rangle$  and
$\langle 0 \mid Q_5\mid \pi\rangle$.  Following \cite{12} consider

\be
\langle 0 \mid j_{\mu 5} \mid \eta \rangle = \tilde{F}q_{\mu}
\ee
$\tilde{F}$ is of order of $f_{\pi}(m_s/M)$ and  can be put to zero in our
approxamation. By taking the divergence from (65), we have

\be
\langle 0 \mid D_s + 6Q_5 \mid \eta \rangle = 0
\ee
The use of (53)  gives (the $\pi-\eta$ mixing as well terms of order
$m_u/m_s,~m_d/m_s$ may be neglected here):

\be
\langle 0 \mid Q_5 \mid \eta \rangle =  \frac{1}{2}
\sqrt{\frac{1}{6}}f_{\pi}m^2_{\eta}
\ee
Relation (67) was found in \cite{12}. By the same reasoning it is easy to
prove that

\be
\langle 0 \mid D_q \mid P_3\rangle + \langle 0 \mid D_s \mid P_3\rangle +
6 \langle 0 \mid Q_5 \mid P_3\rangle =0
\ee
The first term in (68)  is given by (48), the second one is zero according
(57).  For the last term we can write using (43)

\be
\langle 0 \mid Q_5 \mid P_3\rangle = \langle 0 \mid Q_5 \mid \pi \rangle +
\theta \langle 0 \mid Q_5 \mid P_8\rangle
\ee
Eq.'s (68),(69) give

\be
\langle 0 \mid Q_5 \mid \pi \rangle = -\frac{1}{2\sqrt{2}} f_{\pi}m^2_{\pi}
\frac{m_u -m_d}{m_u + m_d}
\ee

-- the same formula as in the case of two light quarks.

It is clear, that the presented above considerations can be generalized to the
case,
 when u,d and s-quark masses are comparable. The calculation became more
cumbersome,
but nothing principially new arises on this case.

\section{$q^2$--dependence of $\chi(q^2)$ at low $q^2$.}

Let us dwell on the  calculation of the $q^2$--dependence of $\chi(q^2)$ at
low $\mid q^2 \mid$ in QCD, restricting ourselves by the first order terms
in the ratio $q^2/M^2$, where $M$ is the characteristic hadronic scale, $M^2
\sim 1~GeV^2$. By $\chi(q^2)$ I mean its      nonperturbative part with
perturbative contribution subtracted from the total $\chi(q^2)$ defined by
(2). The reason is that the perturbative part is strongly divergent, its
contribution would be strongly dependent on regularization proceedure and,
therefore, physically meaningless. In this domain of $q^2~\chi (q^2)$ can be
represented as

\be
\chi(q^2) = \chi(0) + \chi^{\prime}(0)q^2 + R(q^2) - R(0)
\ee
$\chi(0)$ for the QCD case--three light quarks with  $m_u,m_d \ll m_s$ --
was determined in Sec.IV. $\chi^{\prime}(0)$ (its nonperturbative part) for
massless quarks was found in \cite{1}  basing on connection of
$\chi^{\prime}(0)$ with the part of proton spin $\sum$ carried by $u,d,s$
quarks. Its numerical value is given by (4). What is left, is the
contribution of light pseudoscalar quasi--Goldstone bosons $R(q^2)$, which
has nontrivial $q^2$--dependence and must be accounted separately. $R(q^2)$
vanishes for massless quarks and did not contribute to $\chi^{\prime}(0)$,
calculated in \cite{1}. $R(0)$ must be subtracted from $R(q^2)$ since it was
already accounted in $\chi(0)$. $R(q^2)$  can be written as
\be
R(q^2) = - \langle 0 \mid Q_5 \mid \pi \rangle^2  \frac{1}{q^2
-m^2_{\pi}} - \langle 0 \mid Q_5 \mid \eta \rangle^2  \frac{1}{q^2
-m^2_{\eta}}
\ee
The problem of $\eta - \pi$ mixing is irrelevant in the difference $R(q^2) -
R(0)$  in any domain $\mid q^2 \mid \sim m_{\pi}^2$ and $\mid q^2 \mid \sim
m^2_{\eta}$. The matrix elements entering (72)  are given by (67),(70).
Taking the difference $R(q^2) - R(0)$ and using the Euclidean variable $Q^2
=-q^2$, we have

\be
\chi(Q^2) = \chi(0) - \chi^{\prime}(0)Q^2 - \frac{1}{8}f^2_{\pi}Q^2 \Biggl [
\Biggl ( \frac{m_u -m_d}{m_u + m_d} \Biggr )^2
\frac{m^2_{\pi}}{Q^2+m^2_{\pi}} + \frac{1}{3}\frac{m^2_{\eta}}{Q^2
+m^2_{\eta}}\Biggr ]
\ee
Eq.(73) is our final result, where $\chi(0)$ is given by (64) and
$\chi^{\prime}(0)$ by (4). The  accuracy of (73) is given by the parameters
$Q^2/M^2$, $m^2_{\pi}/M^2$, $m^2_{\eta}/M^2 \ll 1$, (two last characterize
the accuracy of $SU(3)\times SU(3)$). At $Q^2\approx m^2_{\eta}$  the last
term comprise about 20\% of the second (the first term is very small,
$\chi(0)\approx -4\cdot 10^{-5}~GeV^4$). Evidently, the last term is much
bigger in the
Minkovski domain, $Q^2 < 0$, since there are pion and $\eta$  poles. As was
mentioned above, $\chi(0)$ found here concides with $\chi(0)$ obtained in
\cite{6} by considering large $N_c~$ limit. However, the $q^2$-dependence is
completely different.
Namely, for $\chi(q^2)$ in \cite{6} was found the relation (eq.($A4^{\prime}$)
in
\cite{6})
\be
\chi(q^2) = - \frac{aF^2_{\pi}}{2N_c} \Biggl [ 1 - \frac{a}{N_c}\sum_i \frac{1}
{q^2 - \mu_i^2} \Biggr ]^{-1},
\ee
where the Goldstone boson masses $\mu_i^2$ are related to quark condensate by
\be
\mu_i^2 = -2m_i \frac{1}{F^2_{\pi}} \langle 0 \mid \bar{q} q \mid 0 \rangle
\ee
and a is some constant of order of hadronic mass square. At $q^2=0$ follows
eq.3 for
$\chi(0)$ if the inequality $a/{N_c\mu_i^2}>>1$ is assumed. However, at
$|q^2|\sim
m^2_{\eta}$ , $m^2_{\pi}$ (74) strongly differs from (73): in (74) there are
zeros at
the points $q^2 = m^2_{\eta}$ , $m^2_{\pi}$, but not poles, as it should be and
as it
take place in (73). And also the most important at low $Q^2$ hadronic term
$\chi^{\prime}(0)Q^2$ is absent in (74).

\section{Summary}

The $q^2$-dependence of topological charge density correlator $\chi(q^2)$ (2)
in QCD
was considered in the domain of low $q^2$. For the cases of two and three light
quarks
the values of  $\chi(0)$, obtained earlier \cite{4}-\cite{7} were rederived
basing on
the low energy theorems and accounting of quasi-Goldstone boson
($\pi$,$\:\eta$)
contributions. No large $N_c$ limit was used and it was no appeal to the
$\theta$-dependence of QCD Lagrangian (except of the proof of absence of
anomalous
commutator in the sum rule (17)).
The only concept, which was used, was the absence of Goldstone boson
contribution as
intermediate state in the singlet (in flavour) axial current correlator in the
limit of
massless quarks.
In the three light quark case -- the case of real QCD -- the mixing of $\pi$
and $\eta$
is of importance and was widely exploited. The $q^2$ dependence of $\chi(q^2)$
was found as arising from two sources:
\begin{itemize}

\item{1}. The contribution of hadronic states (besides $\pi$ and $\eta$). This
contribution was determined from the established in \cite{1} (basing on QCD sum
rule
approach) connection of $\chi^{\prime}(0)$ with the part of the proton spin,
carried by
quarks.
\item{2}. The contributions of $\pi$ and $\eta$ intermediate states. These
contributions
 were calculated by using low energy theorems only.
The final result is presented in eq.73.
\end{itemize}

\newpage

\centerline{\large \bf Acknowledgments.}

\vspace{2mm}
I am thankful to A.V.Smilga for valuable discussions. I am very indebted to
J.Speth for the hospitality at the Institut f\"ur Kernphysik, FZ J\"ulich,
where this work was finished, to A.v.Humboldt Foundation for financial
support of this visit and to V.Baru for a help in the preparation of the
manuscript.
This work was supported in part by CRDF grant
RP2-132,  Schweizerischer National Fonds grant 7SUPJ048716 and RFBR grant
97-02-16131.


\begin{references}
\bibitem{1} B. L. Ioffe and A. G. Oganesian, Phys. Rev. D {\bf 57}, R6590
(1998).
\bibitem{2} A. A. Belyaev, A. M. Polyakov, A. S. Schwartz and
Yu. S. Tyupkin, Phys. Lett.  {\bf 59B}, 85 (1975).
\bibitem{3} G.'t Hooft, Phys. Rev. Lett. {\bf 37}, 8 (1976); Phys. Rev. D
{\bf 14}, 3432 (1976).
\bibitem{4} R. J. Crewther, Phys. Lett. {\bf 70B}, 349 (1977).
\bibitem{5} G. Veneziano, Nucl. Phys  {\bf B159}, 213 (1979).
\bibitem{6} P. Di Vecchia and G. Veneziano, Nucl. Phys. {\bf B171}, 253
(1980).
\bibitem{7}  H. Leutwyler and A. Smilga, Phys. Rev. D {\bf  46},
5607 (1992).
\bibitem{8} B. L. Ioffe, Sov. J. Nucl. Phys. {\bf 29}, 827
(1979).
\bibitem{9} D. J. Gross, S. B. Treiman and F. Wilczek, Phys. Rev. D
{\bf 19}, 2188 (1979).
\bibitem{10} M. Gell-Mann, R. J. Oakes and B. Renner, Phys. Rev.
{\bf 175}, 2195 (1968).
\bibitem{11} B. L. Ioffe and M. A. Shifman,
Phys. Lett. {\bf 95B}, 99 (1980).
\bibitem{12} V. A. Novikov et al., Nucl. Phys. {\bf B165}, 55 (1980).
\end{references}
\end{document}